# PHASE INTEGRAL OF ASTEROIDS

Vasilij G. Shevchenko[1,2], Irina N. Belskaya[2], Olga I. Mikhalchenko[1,2], Karri Muinonen[3,4], Antti Penttilä[3], Maria Gritsevich[3,5], Yuriy G. Shkuratov[2], Ivan G. Slyusarev[1,2], Gorden Videen[6]

[1] Department of Astronomy and Space Informatics, V.N. Karazin Kharkiv National University, Kharkiv 61022, Ukraine

[2] Institute of Astronomy, V.N. Karazin Kharkiv National University, Kharkiv 61022, Ukraine

[3] Department of Physics, University of Helsinki, Gustaf Hällströmin katu 2, FI-00560 Helsinki, Finland

[4] Finnish Geospatial Research Institute FGI, Geodeetinrinne 2, FI-02430 Masala, Finland

[5] Institute of Physics and Technology, Ural Federal University, Mira str. 19. 620002 Ekaterinburg, Russia

[6] Space Science Institute, 4750 Walnut St. Suite 205, Boulder CO 80301, USA

## ABSTRACT

The values of the phase integral $q$ were determined for asteroids using **(i)** a numerical integration of the brightness phase functions over a wide phase-angle range and **(ii)** the relations between $q$ and the $G$ parameter of the $HG$-function and $q$ and the $G_1$, $G_2$ parameters of the $HG_1G_2$-function. The phase-integral values for asteroids of different albedo range from 0.34 to 0.54 with an average value of 0.44. These values can be used for the determination of the Bond albedo of asteroids. Estimates for the phase-integral values using the $G_1$ and $G_2$ parameters are in very good agreement with the available observational data. We recommend using the $HG_1G_2$-function for the determination of the phase integral. Comparison of the phase integrals of asteroids and planetary satellites shows that asteroids have systematically lower values of $q$.

## 1. Introduction

The phase integral $q$ is one of the fundamental characteristics of light scattering by a planetary surface. It is related to the Bond albedo $A$ ($A = pq$, where $p$ is the geometrical albedo; e.g., Shepard, 2017). The Bond albedo is used in the thermal equilibrium equation to model thermal properties of asteroids (e.g., Morrison 1977; Tedesco et al., 2002; Delbo et al., 2003; Masiero et al., 2011; Usui et al., 2011). The phase integral has been defined as:

$$q = 2\int_0^\pi f(\alpha) \sin\alpha \, d\alpha, \qquad (1)$$

where $f(\alpha)$ is the normalized disk-integrated phase function, and $\alpha$ is the solar phase angle. A direct calculation of phase integrals is impossible for the asteroid majority due to the limited phase angle range of ground-based observations. Russell (1916) has estimated the phase integral for the four asteroids (1) Ceres, (2) Pallas, (3) Juno, and (4) Vesta to be equal to 0.55 using an empirical law. Note that they are the largest objects of the asteroid belt. It is not completely clear whether any conclusions concerning these bodies can be simply generalized for the large number of smaller objects, which can have experienced different evolutions. Also according to Russell (1916), the phase integral can be derived from the value of disk-integrated phase function at a phase angle of 50°.

Different researchers have used different estimates of asteroid phase integral for analyzing data obtained in the infrared wavelength range. For example, Morrison (1977) deduced the value of the phase integral $q$ to be equal to 0.6. In analyzing data obtained from the IRAS, WISE, and AKARI satellites, a relationship between $q$ and parameter $G$ of the $HG$-function (Bowell et al., 1989) was used (Tedesco et al. 2002; Masiero et al. 2011; Usui et al., 2011). Since for most asteroids the parameter $G$ was considered to be equal to 0.15, this resulted in a value of the phase integral of 0.384 (Mainzer et al., 2011).

A rigorous determination of the phase integral needs measurements of the disk-integrated phase function over the range of phase angle from 0° to 180°. Unfortunately, ground-based observations allow one to observe main-belt asteroids only from 0° to about 30° phase angles. In this range, the phase functions are well-known for different asteroid taxonomical classes (e.g., Belskaya and Shevchenko, 2000; Harris et al., 1989b; Shevchenko et al., 1997, 2008, 2012, 2015, 2016; Slivan et al., 2008; Slyusarev et al., 2012). Some data were obtained in a wider range of phase angles up to 90° for near-Earth asteroids (Harris et al., 1987; Kaasalainen et al., 2004; Mottola et al., 1997; Hicks et al., 2014, etc.), but such phase functions can be influenced by aspect variations (Muinonen and Wilkman, 2016).

Space missions have allowed various phase functions to be determined up to 160° for different asteroids (Clark et al., 1999; Helfenstein et al., 1994; Masoumzadeh et al., 2015; Newburn et al., 2003; Spjuth et al., 2012, etc.), but the data at small phase angles are often not available. Space-based data complemented with Earth-based data acquired at small phase angles for some asteroids can be used for the numerical calculation of the phase integral. Such data were used by Helfenstein et al. (1994, 1996) for the determination of the phase integral of asteroids (951) Gaspra ($q$=0.47) and (243) Ida ($q$=0.34), by Clark et al. (1999) for (253) Mathilde ($q$=0.28), by Spjuth et al. (2012) for (2867) Steins ($q$=0.59), by Li et al. (2004) for (433) Eros ($q$=0.40), by Masoumzadeh et al. (2015) for (21) Lutetia ($q$=0.40), and by Tatsumi et al. (2018) for (25143) Itokawa ($q$=0.13). However, a detailed analysis of the phase integral for different asteroids, as it was done for planetary satellites (Verbiscer and Veverka, 1988; Brucker et al., 2009), has not yet been performed. Here we investigate how the phase integral depends on the asteroid taxonomical class and compare the phase integrals of asteroids and planetary satellites.

## 2. Average phase functions of brightness in a wide range of phase angles

Detailed observations of the magnitude-phase dependencies of asteroids have revealed their similarity within the main taxonomic classes (Belskaya and Shevchenko, 2000). We construct and present the composite magnitude-phase dependencies for high (E-complex), moderate (S-complex), and low albedo (C-complex) asteroids at small phase angles in Fig. 1. These data were combined from observational data (Belskaya et al., 2003; Binzel et al., 1993; Buchheim, 2010, 2011; Dovgopol et al., 1992; Harris et al., 1984, 1989a, 1989b, 1992; Shevchenko et al., 1996, 2002, 2008, 2010, 2016; Slivan et al., 2008) of magnitude-phase dependencies of individual asteroids (denoted by different symbols in the figures). Data alignment was applied using a shift along the magnitude axis to obtain the best fit between the curves. The alignment of phase curves was carried out by a minimum of the dispersion with linear least squares fit in the overlap region of 10-30 degrees.

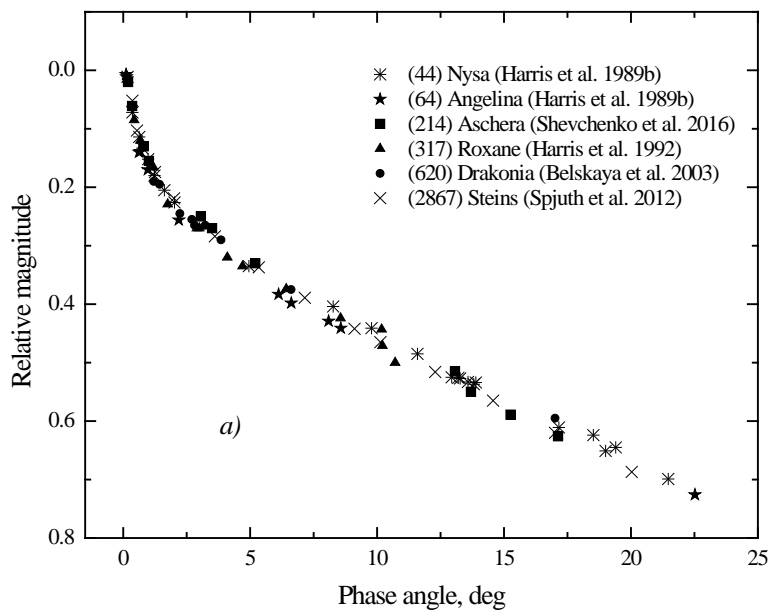

Figure 1a. A composite phase function of brightness for high albedo asteroids ($p$=0.39-0.53) at small phase angles.

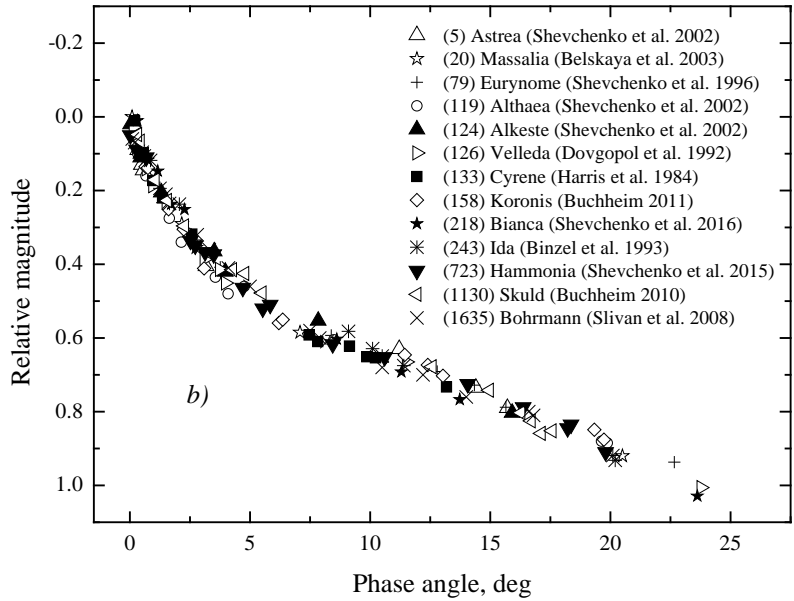

Figure 1b. A composite phase function of brightness for moderate albedo ($p$=0.15-0.25) asteroids at small phase angles.

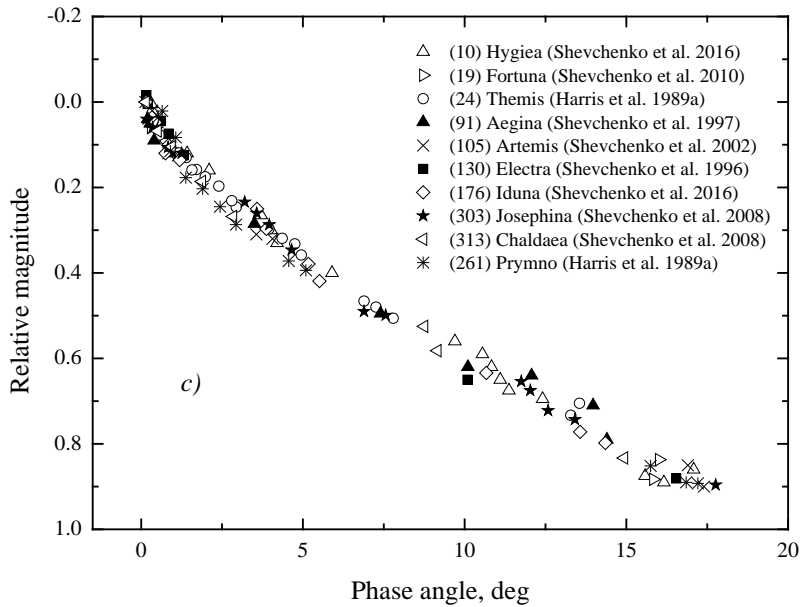

Fig. 1c. A composite phase function of brightness for low albedo ($p$=0.045-0.10) asteroids at small phase angles.

Although the low albedo asteroids show diversity in their brightness phase curves near the opposition, we use for them the average phase function at small α (Shevchenko and Belskaya 2010). The standard deviation of such a composite phase function is about 5%.

To obtain the average phase functions of brightness for asteroids of high, moderate and low albedo over a wide range of phase angles up to 160º, we use the most precise data from space observations (Clark et al., 1999; Masoumzadeh et al., 2015; Newburn et al., 2003; Spjuth et al., 2012). These data were supplemented with data from ground-based observations of some near Earth asteroids (Harris et al., 1987; Kaasalainen et al., 2004; Mottola et al., 1997). Such phase functions of brightness over a wide range of phase angles for selected asteroids are presented in Fig. 2. As one can see, the brightness behavior of asteroids with differing geometric albedo shows differences mainly in the range of the opposition effect and in the range of phase angles of 120–140º.

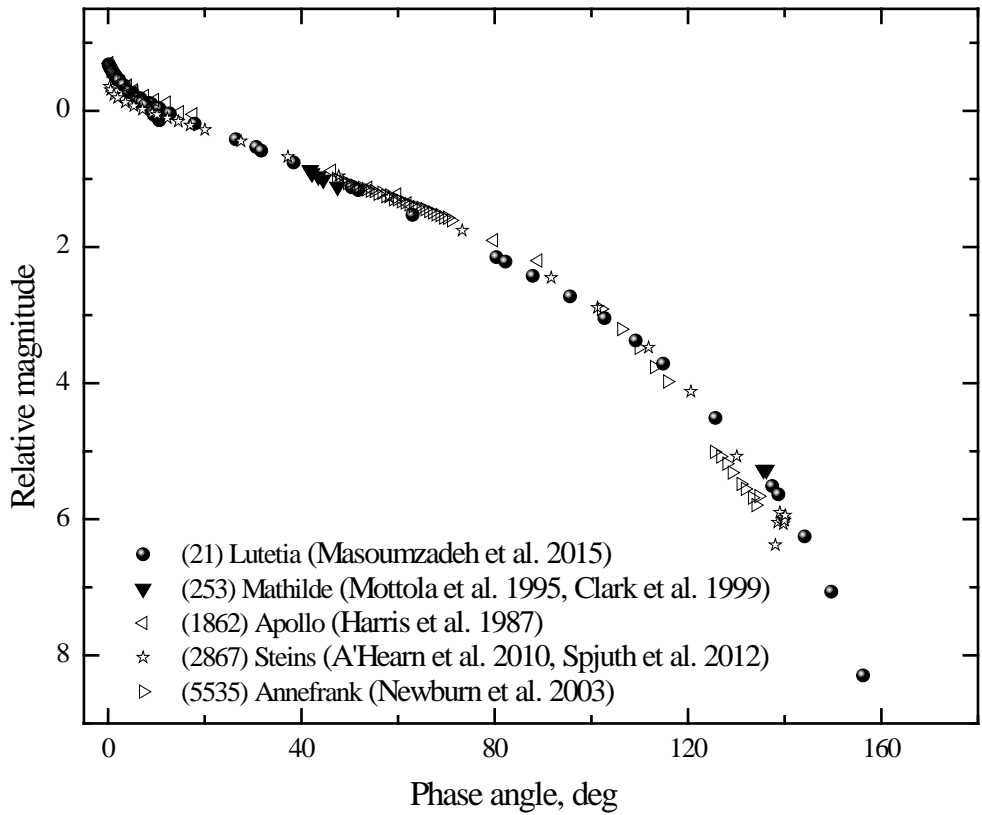

Fig. 2. Phase functions for selected asteroids over a wide range of phase angles: (21) Lutetia, S-complex, $p = 0.19$ (Masoumzadeh et al. 2015); (253) Mathilde, C-complex, $p = 0.047$ (Clark et al. 1999); (1862) Apollo, S-complex, $p = 0.32$ (Nugent et al. 2015); (2867) Steins, E-complex, $p = 0.39$ (Spjuth et al. 2012); (5535) Annefrank, S-complex, $p = 0.24$ (Newburn et al., 2003).

These differences are small (not more than one magnitude) compared to the general brightness variations, but we take into account these differences and use the average phase function for high- (~45%), moderate- (~20%), and low-albedo (~6%) asteroids to obtain a more reliable estimation of the phase integral.

## 3. Phase integrals for asteroids of different albedo

Figure 3 shows the functions $f(\alpha)\sin\alpha$ for moderate-, high-, and low-albedo asteroids, used to calculate the values of the phase integral for these groups. Since the value of the function $f(\alpha)\sin\alpha$ is zero at $\alpha = 180°$, this value was added for all asteroids for numerical calculation of the phase integral. It is also clear from the figure that the behavior of the functions $f(\alpha)\sin\alpha$ is similar in the range of phase angle from 0° to 10°, regardless of the albedo of the surface. It should also be noted that the contribution of the values of the function $f(\alpha)\sin\alpha$ at phase angles >115° in the estimations of the phase integral does not exceed 1%. In addition, the maximum of the function $f(\alpha)\sin\alpha$ depends nonlinearly on the geometrical albedo of the asteroids. This requires an additional study, unfortunately, data on the phase dependencies of brightness for asteroids with albedo >60% are absent. Moreover, the actual existence of asteroids having geometric albedo larger than 0.60 is uncertain. There are a few such objects in the asteroid belt, but their size is less than 15 km, which makes it difficult to obtain high-quality magnitude-phase relations for them.

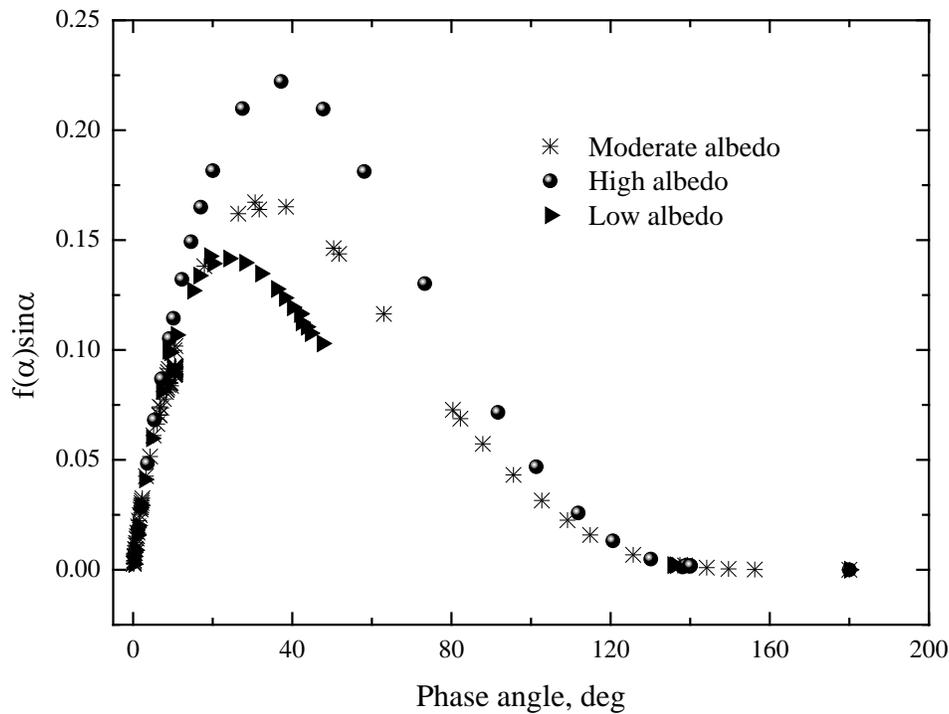

Fig. 3. Function $f(\alpha)\sin\alpha$ for high-, moderate- and low-albedo asteroids.

By numerical integration of $f(\alpha)\sin\alpha$ for asteroids with different geometric albedos, the values of their phase integrals are presented in Table 1. These values lie in the range from 0.35 to 0.54 with the average of 0.44.

The values of the phase integral can also be obtained by integrating the modelled phase functions. For example, the value of the phase integral of the well-known Lommel-Seeliger phase function is equal to 1.64 (e.g., Shepard, 2017), which is much greater than that calculated from the measured phase functions. For real asteroid surfaces, shadowing among the regolith particles/structures causes a steep decrease of the phase curve and thus reduces the value of the phase integral. At present, several phase function models are used in planetary photometry with a number of free parameters for describing the properties of planetary surfaces (e.g., Hapke, 2012, Bowell et al., 1989, Muinonen et al., 2010, Shkuratov et al., 2012, 2018). The often-used function by Hapke (2012) has a large number of parameters and requires both disk-integrated and disk-resolved data to retrieve these parameter values (Clark et al., 1999; Li et al., 2004; Simonelli et al., 1998; Thomas et al., 1996, etc.), but the results can still be ambiguous (Shkuratov et al., 2012). A new function proposed by Shkuratov et al. (2018) has a small number of parameters and works well for different classes of objects for both disk-integrated and disk-resolved data, but currently these parameters have not been estimated for asteroids of different albedos, and the connection between the phase integral and these parameters has not been studied. Here we use the *HG* and $HG_1G_2$ functions (Bowell et al., 1989, Muinonen et al., 2010), which were recommended by the IAU as magnitude systems in asteroid integral photometry. There are relationships of the phase integral with parameter *G* (Bowell et al., 1989):

$$q = 0.290 + 0.684G, \qquad (2)$$

and $G_1$ and $G_2$ (Muinonen et al., 2010):

$$q = 0.009082 + 0.4061G_1 + 0.8092G_2, \qquad (3)$$

for these functions. Typical dispersion of the *G* parameter among asteroids is 0.10, and 0.05 for $G_1$ and $G_2$. Using the average parameters of *G*, $G_1$ and $G_2$ for the main taxonomical classes (Shevchenko and Lupishko, 1998; Shevchenko et al., 2003; Shevchenko et al., 2016) and the relations between *q* and these parameters, the average values of the phase integral for asteroids with different albedo surfaces were obtained. These data are listed in Table 1 where the values of the phase integral calculated with formulas (1)–(3) are presented.

The average values of the phase integral calculated from the observations and from the *HG* and $HG_1G_2$ functions coincide well, although the value of the phase integral for high-albedo asteroids using the *HG* model has a significantly greater value. However, it should be noted that the *HG*-function has a reduced brightness in the area of the opposition region relative to real behavior, in contrast to the $HG_1G_2$-function, which more accurately approximates the phase dependency of brightness. Thus, the $HG_1G_2$-function produces results that are more accurate when brightness values are only available in the range of phase angles from 0° to 30°.

The values of phase integral for moderate-albedo asteroids are close to the values obtained by Helfenstein et al. (1994) for the asteroid (951) Gaspra (*q* = 0.47), by Li et al. (2004) for (433) Eros (0.40), by Masoumzadeh et al., (2015) for (21) Lutetia (0.40), and by Hicks et al. (2014) for (4) Vesta (0.44), though this asteroid has an albedo of about 40%. The values of the phase integral for low-albedo asteroids obtained in this work differ significantly from the values obtained by Clark et al. (1999) for the low-albedo asteroid (253) Mathilde ($q = 0.28$). In our opinion, this is due to the inadequate combination of phase dependencies derived from Earth-based observations (Mottola et al., 1995) and spacecraft data for this asteroid. We obtained also value of the phase integral for (1) Ceres using data for phase function of brightness from Ciarniello et al. (2017) and Tedesco et al. (1983). The value is equal to $0.35 \pm 0.02$ and is typical for low albedo asteroids.

Table 1. The values of the phase integral for asteroids of differing geometric albedo. The phase integral $q$ for the observations was computed with numerical integration. The error bars are the 1-sigma dispersion in the calculated values.

|  | Geometric albedo | Phase integral $q$ | | |
| --- | --- | --- | --- | --- |
|  |  | Observations | $HG$ function | $HG_1G_2$ function |
| Low albedo | $0.061 \pm 0.017$ | $0.35 \pm 0.02$ | $0.34 \pm 0.03$ | $0.36 \pm 0.02$ |
| Moderate albedo | $0.20 \pm 0.05$ | $0.42 \pm 0.02$ | $0.45 \pm 0.02$ | $0.42 \pm 0.02$ |
| High albedo | $0.45 \pm 0.07$ | $0.54 \pm 0.02$ | $0.63 \pm 0.04$ | $0.56 \pm 0.03$ |
| Average | $0.24 \pm 0.20$ | $0.44 \pm 0.10$ | $0.47 \pm 0.15$ | $0.45 \pm 0.10$ |

## 4. Comparison with planetary satellites

Using the values of the parameters $G_1$ and $G_2$ (Shevchenko et al., 2016; Penttilä et al., 2016) for about one hundred different-type asteroids, the values of the phase integral were determined and compared with those obtained by Brucker et al., (2009) for planetary satellites. In addition, the phase dependencies of brightness for the Moon, Phobos and Deimos (Avanesov et al., 1991; Bowell et al., 1989; Rougier, 1933; Velikodsky et al., 2011) were used for the determination of their phase integrals, which are $0.48 \pm 0.02$, $0.38 \pm 0.03$, $0.40 \pm 0.03$, respectively. Our value of the phase integral for Phobos ($0.38 \pm 0.03$) differs from the value of $0.30 \pm 0.04$ obtained by Simonelli et al. (1998), although the value for Deimos ($0.40 \pm 0.03$) is in agreement with the value of $0.39 \pm 0.02$ obtained by Thomas et al. (1996). In the case of the Moon, our value is very different from the value 0.60 obtained by Lane and Irvin (1973). We assume that it is related to the use of new, better quality data on the phase function of the Moon obtained by Velikodsky et al. (2011). Figure 4 shows the dependency of the phase integrals for asteroids and planetary satellites as a function of their geometric albedos. It should be noted that the albedo range is wider for the satellites and there are currently no data on the phase integrals for asteroids with albedos greater than 60%.

In Fig. 4, the solid line shows a linear fit for the satellites without Phoebe and Europa, and a dot-dashed line shows a linear fit when Phoebe and Europa are included. We added data for the Moon, Phobos, Deimos, and asteroids on the satellite diagram without a recalculation of the linear fits for the satellites given by Brucker et al. (2009). The dashed line shows a linear fit only for the asteroids ($q = 0.359$ (±0.005) + 0.47 (±0.03) $p$). In general, the values of phase integrals of satellites are systematically larger than for asteroids. This indicates that the linear part of phase functions of the asteroids has a greater slope than for the satellites. This result remains to be interpreted from a theoretical point of view. An exception is the Saturnian satellite Phoebe that has the lowest value of the phase integral among the objects under study. The Martian satellites Deimos and Phobos, and Uranian satellite Miranda have phase-integral values similar to those of asteroids. When extrapolating the asteroid phase integrals to albedos greater than 60 %, we retrieve values similar to those of the Saturnian satellites of Rhea, Tethys, and Enceladus, i.e., not greater than 0.8. If this extrapolation holds true, then we can expect that for such asteroids, the phase function differs from the behavior of E-type asteroids and their thermal properties may not correspond to the developed thermal models for high-albedo asteroids.

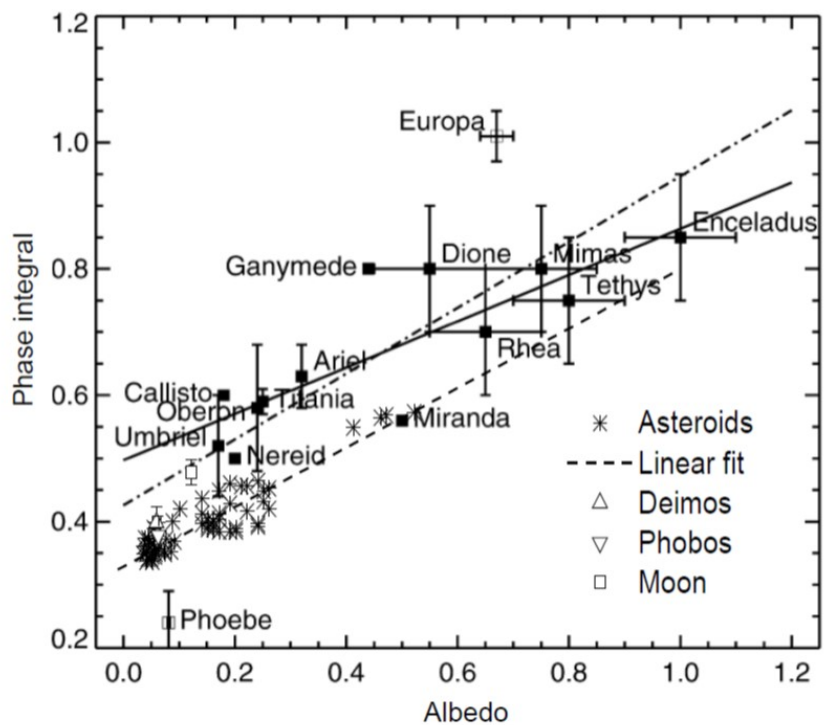

Fig. 4. Dependency of the phase integral on the geometric albedo for asteroids and planetary satellites, the dashed line is the best linear fit to the asteroid data. Adapted from Brucker et al. (2009). The solid line is the adopted best linear fit to the satellite data without including Phoebe and Europa and the dot-dashed line is the best linear fit to the data including Phoebe and Europa.

## 5. Conclusion

Based on the results of ground-based observations and data obtained from space missions, the composite average phase dependencies of brightness for moderate, high and low albedo asteroids in a wide range of phase angles from 0º to 160º were obtained. Because the phase functions of brightness for asteroids inside a taxonomic class behave similarly, it is possible to estimate the value of the phase integral for that asteroid class using a representative phase function.

The values of the phase integral $q$ were determined for asteroids of different albedo using both (i) a numerical integration of the average brightness phase functions over a wide phase-angle range and (ii) the relations between $q$ and the $G$ parameter of the $HG$-function and between $q$ and the $G_1$, $G_2$ parameters of the $HG_1G_2$-function. The values of the phase integral lie in the range from 0.34 to 0.63 with an average of 0.45. These data can be used to model the thermal properties of asteroid surfaces and to process data obtained in the infrared wavelength range.

The behavior of the function $f(\alpha)\sin\alpha$ is similar in the range of phase angle from 0º to 10º, regardless of the surface albedo. Moreover the contribution of the function $f(\alpha)\sin\alpha$ at phase angles greater than 115º using an estimation of the phase function does not exceed one per cent, and has a limited impact on the value of the phase integral calculated. In addition, the maximum of the function $f(\alpha)\sin\alpha$ depends nonlinearly on the asteroid albedo. Such behavior warrants additional study, but, unfortunately, data on phase dependencies of brightness for asteroids with albedos greater than 60% are absent. An estimation of the phase integral values using $G_1$ and $G_2$ parameters gives a very good agreement with data obtained from space missions. We recommend using the $HG_1G_2$-function for the determination of the phase integral. In the case of unknown phase function for an asteroid, it makes sense to use the average value of the phase integral for an asteroid of corresponding albedo and/or the linear dependency on albedo. The differences in the values of the phase integral for asteroids of different classes are important to take into account in thermal modeling. A comparison of the phase integrals shows that asteroids have systematically lower values than planetary satellites having the same albedo. Moreover, when asteroid phase integrals are extrapolated into regions of greater albedo, their values are less than 0.8. As mentioned above the actual existence of asteroids having geometric albedo larger than 0.60 is uncertain. It can be expected that, for asteroids, the phase integral cannot exceed the value of 0.8.

*Acknowledgments.* We are grateful to Dr. A. Cellino for constructive comments that improved our article. VGS thanks the University of Helsinki for supporting the participation in European Planetary Science Conference. This research was supported by the Ukrainian Ministry of Education and Science, by Kharkiv regional fond named after K. D. Sinelnikov and in part by the ERC Advanced Grant No. 320773 (K. Muinonen, M. Gritsevich).

# References

A'Hearn, M. F., Feaga, L. M., Bertaux, J.-L., et al. 2010, Planet. Space Sci., 58, 1088
Avanesov, G., Zhukov, B., Ziman, Ya., et al. 1991, Planet. Space Sci., 39, 281
Belskaya, I. N. & Shevchenko, V. G. 2000, Icarus, 147, 94
Belskaya, I. N., Shevchenko, V. G., Kiselev, N. N., et al. 2003, Icarus, 166, 276
Binzel, R. P., Slivan, S. M., Magnusson, P., et al. 1993, Icarus, 105, 310
Bowell, E., Hapke, B., Domingue, D., et al. 1989, in Asteroids II, eds. Binzel, R.P., Gehrels, T. & Matthews, M.S. (Tucson: University of Arizona Press), 524–556
Brucker, M. J., Grundy, W. M., Stansberry, J. A., et al. 2009, Icarus, 201, 284
Buchheim, R. K. 2010, Minor Planet Bull., 37, 41.
Buchheim, R. K. 2011, Minor Planet Bull., 38, 128.
Ciarniello, M., De Sanctis, M. C., Ammannito, E., et al. 2017, A&A, 598, A130.
Clark, B. E., Veverka, J., Helfenstein, P., et al. 1999, Icarus, 140, 53
Delbó, M., Harris, A. W., Binzel, R. P., Pravec, P., Davies, J. K. 2003, Icarus, 166, 116
Dovgopol, A. N., Krugly, Yu. N., Shevchenko, V. G. 1992, Acta Astron., 42, 67
Hapke, B., 2012. Theory of Reflectance and Emittance Spectroscopy, second ed. Cambridge Univ. Press, Cambridge, UK
Harris, A. W, Carlsson, M., Young, J. W., Lagerkvist, C.-I. 1984, Icarus, 58, 377
Harris, A. W., Young, J. W., Goguen, J., et al. 1987, Icarus, 70, 246
Harris, A. W., Young, J. W., Bowell, E., et al. 1989a, Icarus, 77, 171.
Harris, A. W., Young, J. W., Contreiras, L., et al. 1989b, Icarus, 81, 365
Harris, A. W., Young, J. W., Dockweiler, T., et al. 1992, Icarus, 95, 115
Helfenstein, P., Veverka, J., Thomas, P. C., et al. 1994, Icarus, 107, 37
Helfenstein, P., Veverka, J., Thomas, P. C., et al. 1996, Icarus, 120, 48
Hicks, M. D., Buratti, B. J., Lawrence, K. J., et al. 2014, Icarus, 235, 60
Kaasalainen, M., Pravec, P., Krugly, Yu. N., et al. 2004, Icarus, 167, 178
Lane, A. P. & Irvine, W. M. 1973, AJ, 78, 267
Li, J., A'Hearn, M. F., McFadden, L. A. 2004, Icarus, 172, 415
Mainzer, A., Grav, T., Masiero, J., et al. 2011, ApJ, 741, 90
Masiero, J. R., Mainzer, A. K., Grav, T., et al. 2011, ApJ, 741, 68
Masoumzadeh, N., Boehnhardt, H., Li, J.-Y., Vincent, J.-B. 2015, Icarus, 257, 239
Morrison, D. 1977, Icarus, 31, 185
Mottola, S., Sears, W. D., Erikson, A., et al. 1995, Planet. Space Sci., 43, 1609
Mottola, S., Erikson, A., Harris, A. W., et al. 1997, AJ, 114, 1234
Muinonen, K., & Wilkman, O. 2016, in Proceedings of the International Astronomical Union, Symposium S318, Asteroids: New Observations, New Models, eds. S. Chesley, A. Morbidelli, R. Jedicke, Cambridge University Press, 6 p
Muinonen, K., Belskaya, I. N., Cellino, A., et al. 2010, Icarus, 209, 542
Newburn, R. L. Jr., Duxbury, T. C., Hanner, M., et al. 2003, J. Geophys. Res., 108, 5117
Penttilä, A., Shevchenko, V. G., Wilkman, O., Muinonen, K. 2016, Planet. Space Sci., 123, 117
Rougier, G. 1933, Ann. Obs. Strasbourg, 2, 1
Russell, H. N. 1916, ApJ, 43, 173
Shepard, M. K., 2017. Introduction to planetary photometry. Cambridge University Press, Cambridge, United Kingdom, 249 p
Shevchenko, V. G. & Lupishko, D. F. 1998, Solar System. Research, 32, 220
Shevchenko, V. G. & Belskaya, I. N. 2010, in European Planetary Science Congress 2010, 738
Shevchenko, V. G., Chiorny, V. G., Kalashnikov, A. V., et al. 1996, A&A Suppl. Ser., 115, 475

Shevchenko, V. G., Belskaya, I. N., Chiorny V. G., et al. 1997, Planet. Space Sci., 45, 1615
Shevchenko, V. G., Belskaya, I. N., Krugly, Yu. N., et al. 2002, Icarus, 155, 365
Shevchenko, V. G., Krugly, Yu. N., Chiorny, V. G., et al. 2003, Planet. Space Sci., 51, 525
Shevchenko, V. G., Chiorny, V. G., Gaftonyuk, N. M., et al., 2008, Icarus, 196, 601
Shevchenko, V. G., Belskaya, I. N., Lupishko, D. F., et al. 2010. EAR-A-COMPIL-3-MAGPHASE-V1.0. NASA Planetary Data System.
Shevchenko, V. G., Belskaya, I. N., Slyusarev, I. G., et al. 2012, Icarus, 217, 202
Shevchenko, V. G., Slyusarev, I. G., Belskaya, I. N., et al. 2015, in Lunar and Planetary Science Conference, Vol., 46, 1509
Shevchenko, V. G., Belskaya, I. N., Muinonen, K., et al. 2016, Planet. Space Sci., 123, 101
Shkuratov, Y., Kaydash, V., Korokhin, V., et al. 2012, J. Quant. Spectr. Rad. Transf., 113, 2431
Shkuratov, Yu. G., Korokhin, V. V., Shevchenko, V. G., et al. 2018, Icarus, 302, 213
Simonelli, D. P., Wisz, M., Switala, A., et al. 1998, Icarus, 131, 52
Slivan, S. M., Binzel, R. P., Boroumand, S. C., et al. 2008, Icarus, 195, 226
Slyusarev, I. G., Shevchenko, V. G., Belskaya, I. N., et al. 2012, in Lunar and Planetary Science Conference, Vol., 43, 1885
Spjuth, S., Jorda, L., Lamy, P. L., et al. 2012, Icarus, 221, 1101
Tatsumi, E., Domingue, D., Hirata, N., et al. 2018, Icarus, 311, 175
Tedesco, E. F., Taylor, R. C., Drummond, J., et al. 1983, Icarus, 54, 23
Tedesco, E. F., Noah, P. V., Noah, M., Price, S. D. 2002, AJ, 123, 1056
Thomas, P. C., Adinolfi, D., Helfenstein, P., et al. 1996, Icarus, 123, 536
Usui, F., Kuroda, D., Muller, T. G., et al. 2011, Publ. Astron. Soc. Japan, 63, 1117
Velikodsky, Yu. I., Opanasenko, N. V., Akimov, L. A., et al. 2011, Icarus, 214. 30
Verbiscer, A. J. & Veverka, J. 1988, Icarus, 73, 324